\documentclass[a4paper,10pt]{article}

\usepackage{rsfs}
\usepackage{latexsym}
\usepackage{amssymb}
\usepackage{graphicx}
\makeatletter
  
  \@addtoreset{equation}{section}
\makeatother
\date{}
  \title{Effective Coupling Constant in Renormalization Group for the 
  Quantum Electrodynamics}

\author{Hirohisa Ishikawa 
\footnote{E-mail address: ishikawa@meikai.ac.jp}\\
{\small \it Department of Economy, Meikai University,
 Urayasu, Chiba, 278-8550, Japan.}\\
{}
\\
and\\
Keiji Watanabe
\footnote{Present address; 5-36-2 Akazutumi, Setagaya-ku 156-0044}\\
{\small \it Department of Physics, Meisei University,}\\
{\small \it Hino Tokyo, 191-8506, Japan}}
\begin{document}

\maketitle

\begin{abstract}
Effective coupling constant in quantum electrodynamics is investigated. A pole 
appears in the effective coupling constant for the space-like momentum if it is calculated by perturbation. The pole can be eliminated by the analytic regularization. For QED the effective coupling constant is 
written in terms of the scale parameter, $\Lambda$,  having the dimension of 
mass 
as in the case of QCD. 
$\Lambda$ is determined by comparing with the experimental data. The 
calculated result agrees with experiment with $\Lambda\approx 1.64\times 10^{47}$ GeV; it is very large but much smaller than the mass scale of Landau ghost. 
. 
\end{abstract}

\section{Introduction}

The effective coupling constant, $\alpha_S$, is one of the most fundamental
 quantities in the renormalization group. It is a renormalization group invariant quantity so that it is usually used to express 
the Green function. $\alpha_S$ is given in terms of the
 beta function, $\beta$, the derivative of the renormalized coupling constant
  with respect to the renormalization point. Actually, it is defined 
 implicitly by the equation
\begin{equation}
 \log\left(\frac{Q^2}{\mu^2}\right)=\int_{\alpha_r}^{\alpha_S(Q^2)}
 \frac{1}{\beta(\alpha')}d\alpha'.   \label{alpha}
\end{equation}
Here, $Q^2$ is the squared momentum for the space-like region, $\mu$ is the 
renormalization point and the renormalized coupling constant $\alpha_r$ is
given as $\alpha_S(\mu^2)=\alpha_r$.
In this paper we restrict ourselves to the abelian and non-abelian gauge 
theories, the quantum electrodynamics (QED) and the quantum chromodynamics 
(QCD). 

The beta functions are calculated by perturbation with recourse to the 
$\overline{\rm MS}$ 
renormalization scheme.
 For QED, Gorishny et al. \cite{gorishny} performed calculation up to the 
 fourth order approximation and obtained
\begin{equation}
 \beta_{QED}(\alpha)/8\pi=\beta_0(\alpha/4\pi)^2+\beta_1(\alpha/4\pi)^3
 +\beta_2(\alpha/4\pi)^4+\beta_3(\alpha/4\pi)^5+\cdots, \label{qed}
\end{equation}
where $\beta_k, \,k=1,2,\cdots,4$ are
\begin{eqnarray} 
  \beta_0 &=& 2n_f/3\pi,  \nonumber \\
  \beta_1 &=& n_f/2\pi^2, \nonumber \\
  \beta_3 &=& -(1+22n_f/9)n_f/16\pi^3, \nonumber \\
  \beta_4 &=& -n_f\Big[23-(380/27-416\zeta(3))n_f/9+616n_f^2/243\Big],
\end{eqnarray}
with $n_f$ being the number of flavor and $\zeta(n)$ the Riemann $\zeta$ 
function. For QCD, the gauge group being $SU(3)$ with the number of 
flavor $n_f=3$, the three
 loop calculation was done by Tarasov et al. \cite{tarasov}. Their result is 
\begin{equation}
 \beta_{QCD}/4\pi = \beta_0(\alpha/4\pi)^3+\beta_1(\alpha/4\pi)^5
       +\beta_2(\alpha/4\pi)^7+\cdots,        \label{qcd}
\end{equation}
where
\begin{eqnarray}
 \beta_0 &=& (-11+2n_f/3),  \nonumber \\
 \beta_1 &=& -102+38n_f/3,                   \nonumber \\
 \beta_2 &=& -2857/2+25033n_f/18-325n_f^2/54.
\end{eqnarray}

One of the most important results in the renormalization group is that the 
qualitative feature of the effective coupling constant is determined by the 
sign of the lowest order term in the perturbation expansion, namely, the sign of $\beta_0$.
When $\beta_0<0$, the theory is asymptotically free and the $\alpha(Q^2) \to 0$ for  $Q^2 \to \infty$. For $\beta_0>0$ the theory is asymptotically non-free.
 QCD with $n_f<33/2$ is asymptotically free theory, but QED is 
not asymptotically free. 

\section{Analytic regularization}

To simplify the problem, we start by studying the one loop approximation. 
Taking the lowest
 order term in the $\beta$ function, we obtain the effective 
coupling constant: For QED, the integration of (\ref{alpha}) 
 leads to
$$
 \log(Q^2/\mu^2)=
  \frac{2\pi}{\beta_0}\left(\frac{1}{\alpha_r}-\frac{1}{\alpha_S(Q^2)}\right).
$$
Writing the renormalized coupling constant $\alpha_r$ in terms of a parameter 
 $\Lambda$, which is defined by the equation
\begin{equation}
  \Lambda^2=\mu^2\exp\left(\frac{2\pi}{\beta_0\alpha_r}\right), 
\end{equation}
we have
\begin{equation}
 \alpha_S(Q^2)^{QED}=\frac{2\pi}{\beta_0}\frac{1}{\log(\Lambda^2/Q^2)}.
          \label{QED oneloop}
\end{equation}
As $\beta_0>0$ for QED, the effective coupling constant $\alpha_S(Q^2)$ is an 
increasing function of $Q^2$ for $Q^2<\Lambda^2$. $\alpha_S(Q^2)$ is 
singular at $Q^2=\Lambda^2$ and becomes unphysical if $Q^2>\Lambda^2$ because  
$\alpha_S(Q^2)$ turns out to be negative.@For QCD, $Q^2$ dependence is different form QED as $\beta_0(=-\beta'_0)<0$.
For the one loop approximation the scale parameter $\Lambda$ is given as
\begin{equation}
\Lambda^2=\mu^2\exp\left(-\frac{4\pi}{\beta_0'\alpha_r}\right)
\end{equation}
and
\begin{equation}
 \alpha_S(Q^2)^{QCD}=\frac{4\pi}{\beta_0'}\frac{1}{\log(Q^2/\Lambda^2)}. 
                   \label{QCD oneloop}
\end{equation}
It is a decreasing function of $Q^2$ and has a pole at $Q^2=\Lambda^2$. It 
becomes negative and unphysical for $Q^2<\Lambda^2$.

To eliminate the singularity of the effective coupling constant in gauge theories we have adopted the analytic regularization \cite{dokshitzer}. The regularization is performed in the following way: First calculate the effective coupling constant for the space-like region as a function of the squared momentum $Q^2$, 
which is transformed to the time-like squared momentum $t$ by the replacement
\begin{equation}
  Q^2 \to e^{-i\pi}t.
\end{equation}
 For QCD the regularized effective 
coupling constant $\alpha_R$ is defined by the dispersion integral
\begin{equation}
  \alpha_R^{QCD}(t)=\frac{1}{\pi}\int_0^{\infty}\frac{\sigma(t')}{t'-t}dt',
                        \label{QCD R}
\end{equation}
where
\begin{equation}
 \sigma(t)=\frac{1}{2i}[\alpha_S(e^{-i\pi}t)-\alpha_S(e^{i\pi}t)].
\end{equation}
Therefore, the spectral function of QCD is given for the one loop 
approximation as
\begin{equation}
 \sigma(t)=\frac{4\pi}{\beta'_0}
  \frac{1}{\left(\log(t/\Lambda^2)\right)+\pi^2},
\end{equation}
which is a decreasing function of $t$.
Performing integration of (\ref{QCD R}) by means of the Cauchy theorem,
we have
\begin{equation}
 \alpha^{QCD}_R(Q^2)=\alpha_S^{QCD}(Q^2)-\frac{4\pi}{\beta_0}
  \frac{\Lambda^2}{Q^2-\Lambda^2}.
\end{equation}

For QED the spectral function $\sigma$ turns out to be negative, and 
an increasing function of $t$
\begin{equation}
 \sigma(t)=-\frac{2\pi}{\beta_0}\frac{1}{(\log(\Lambda^2/t))^2+\pi^2}.
\end{equation}
Consequently, the regularization should be changed from that of QCD. We take 
once subtracted dispersion formula
\begin{equation}
 \alpha_R^{QED}(t)=\frac{t}{\pi}
   \int_0^{\infty}\frac{\sigma(t')}{t'(t'-t)}dt'. \label{QED R}
\end{equation}
We have the following regularized effective coupling constant for the one loop 
approximation:
\begin{equation}
 \alpha_R^{QED}=\alpha_S^{QED}(Q^2)-\frac{2\pi}{\beta_0}\frac{Q^2}{\Lambda^2-Q^2}.
\end{equation}

Although we have considered one loop calculations so far, we may use 
 (\ref{QCD R}) and (\ref{QED R}) for general case.
 
 For the case of QCD, the properties of effective coupling constant is 
 investigated in \cite{NW}. So, we restrict ourselves to QED in this paper.
 
 Here we use the effective coupling constant for the four loop calculation
  \cite{gorishny}. 
To make the formula simpler, we write the $\beta$ function for QED
\begin{equation}
 \beta(\alpha) = \alpha^2(a_1+a_2\alpha+a_3\alpha^2+a_4\alpha^3+\cdots).
\end{equation}
Calculating (\ref{alpha}), we obtain
\begin{eqnarray}
 -\frac{a_1}{2}\log\left(\frac{Q^2}{\mu^2}\right) &=& \frac{1}{\alpha_S(Q^2)}
  -\frac{1}{\alpha_r}+\frac{a_2}{a_1}
  \log\left(\frac{\alpha_S(Q^2)}{\alpha_r}\right)
  -\frac{1}{a_1^2}(a_2^2-a_1a_3)\left(\alpha_S(Q^2)-\alpha_r\right) 
                                                     \nonumber \\
  &&+\frac{1}{2a_3}(a_2^3-2a_1a_2a_3+a_1^2a_4)\left(\alpha_S(Q^2)
  -\alpha_r^2\right)+\cdots.
\end{eqnarray}
We eliminate the renormalized coupling constant $\alpha_r$ and the
renormalization point $\mu$ by using 
the parameter with dimension of mass, $\Lambda^{QED}$, which is given by 
the equation
\begin{eqnarray}
 \frac{a_1}{2}\log\left[\frac{(\Lambda^{QED})^2}{\mu^2}\right]
  &=& \frac{1}{\alpha_r}-\frac{1}{a_1^2}(a_2^2-a_1a_3)\alpha_r 
  +\frac{1}{2a_1^3}(a_2^3-2a_1a_2a_3+a_1^2a_4)\alpha_r^2+\cdots.
\end{eqnarray}
We simply write $\Lambda$ by omitting the superscript \lq$ QED$ \rq 
 $\,$ hereafter.
The effective coupling constant is now given as follows:
\begin{eqnarray}
 \frac{1}{\alpha_S(Q^2)} &=& \frac{a_1}{2}\log(\Lambda^2/Q^2)
  +\frac{a_2}{a_1}\log\{\log(\Lambda^2/Q^2)\}  \nonumber \\
  &&+\frac{2}{a_1^3\log(\Lambda^2/Q^2)}[a_2^2\log\{\log(\Lambda^2/Q^2)\}
  +a_2^2-a_1a_3]   \nonumber \\
  &&-\frac{2}{a_1^5\log^2(\Lambda^2/Q^2)}
   [a_2^3\log^2\{\log(\Lambda^2/Q^2)\}   \nonumber \\
  &&-2a_1a_2a_3\log\{\log(\Lambda^2/Q^2)\}
  -a_2^3+a_1^2a_4]+\cdots.   \label{alpha(Q^2)}
\end{eqnarray}

\section{Effective coupling constant of QED for the time-like momentum}

We perform the analytic continuation of the effective coupling constant for the four loop calculation (\ref{alpha(Q^2)}) by using the prescription given in
 the previous section. The squared momentum for the 
space-like region, $Q^2$,  is transformed to the time-like one $t$ by the
 equation
$$
 Q^2=e^{-i\pi}t.
$$
(\ref{alpha(Q^2)}) then becomes
\begin{equation}
 1/\alpha_S(e^{-i\pi}t)=u+iv,   \label{time}
\end{equation}
where $u$ and $v$ are given as follows:
\begin{eqnarray}
  u &=& \frac{a_1}{2}R\cos \theta+\frac{a_2}{a_1}\log R 
    +\frac{2}{a_1^3R}[\cos \theta(a_2^2\log R+a_2^2-a_1a_3)
      +\theta a_2^2\sin \theta]  \nonumber \\
    &&-\frac{2}{a_1^5R}[\cos 2\theta \{a_2^3(\log^2 R -\theta^2)
      -2a_1a_2a_3\log R-a_2^3+a_1^2a_4\} \nonumber \\
    &&+2\theta\sin 2\theta(a_2^3\log R-a_1a_2a_3)]  \nonumber \\
 v &=& \frac{a_1}{2}R\sin \theta +\theta \frac{a_2}{a_1}
   +\frac{2}{a_1^3R}[-(a_2^2\log R+a_2^2-a_1a_3)\sin \theta
     +\theta a_2^2\cos \theta] \nonumber \\
   &&-\frac{2}{a_1^5R}[-\sin 2\theta \{a_2^3(\log^2 R-\theta^2)
     -2a_1a_2a_3\log R-a_2^3+a_1^2a_4\} \nonumber \\
   &&-\frac{2}{a_1^5 R}[-\sin 2 \theta \{a_2^3(\log^2 R-\theta^2)
     -2a_1a_2a_3\log R-a_2^3+a_1^2a_4\}  \nonumber \\
   &&+2\theta \cos 2\theta (a_2^3\log R-a_1a_2a_3)],
\end{eqnarray}
with
\begin{equation}
 R=\sqrt{\log^2 (\Lambda^2/t)+\pi^2}
\end{equation}
and
\begin{equation}
 \theta=\arctan\left(\pi/\log (\Lambda^2/t)\right)
\end{equation}
Here the branch of $\arctan\left(\pi/\log (\Lambda^2/t)\right)$ is taken to 
be continuous at $t=\Lambda^2$.
${\rm\, Re}\, \alpha$ and ${\rm\, Im}\, \alpha$ are given as
\begin{eqnarray}
 {\rm Re}\, \alpha(t) &=& \frac{u}{u^2+v^2}, \\
 {\rm Im}\, \alpha(t) &=& -\frac{v}{u^2+v^2}.
\end{eqnarray}
We illustrate in Fig.1 (a), (b) ${\rm Re}\, \alpha(t)$ and ${\rm Im}\, \alpha(t)$ as 
functions of $t$ for the time-like momentum, $t>0$, respectively. Here, we use 
$\Lambda$ which will be obtained by comparing with the experimental data in 
Sec.4. It must be remarked that the spectral 
function $\sigma(t)$ is given by $\sigma(t)={\rm\, Im}\alpha(t)$.
\begin{figure}
\begin{center} 
\begin{tabular}{cc}
  \includegraphics[width=.47\linewidth]{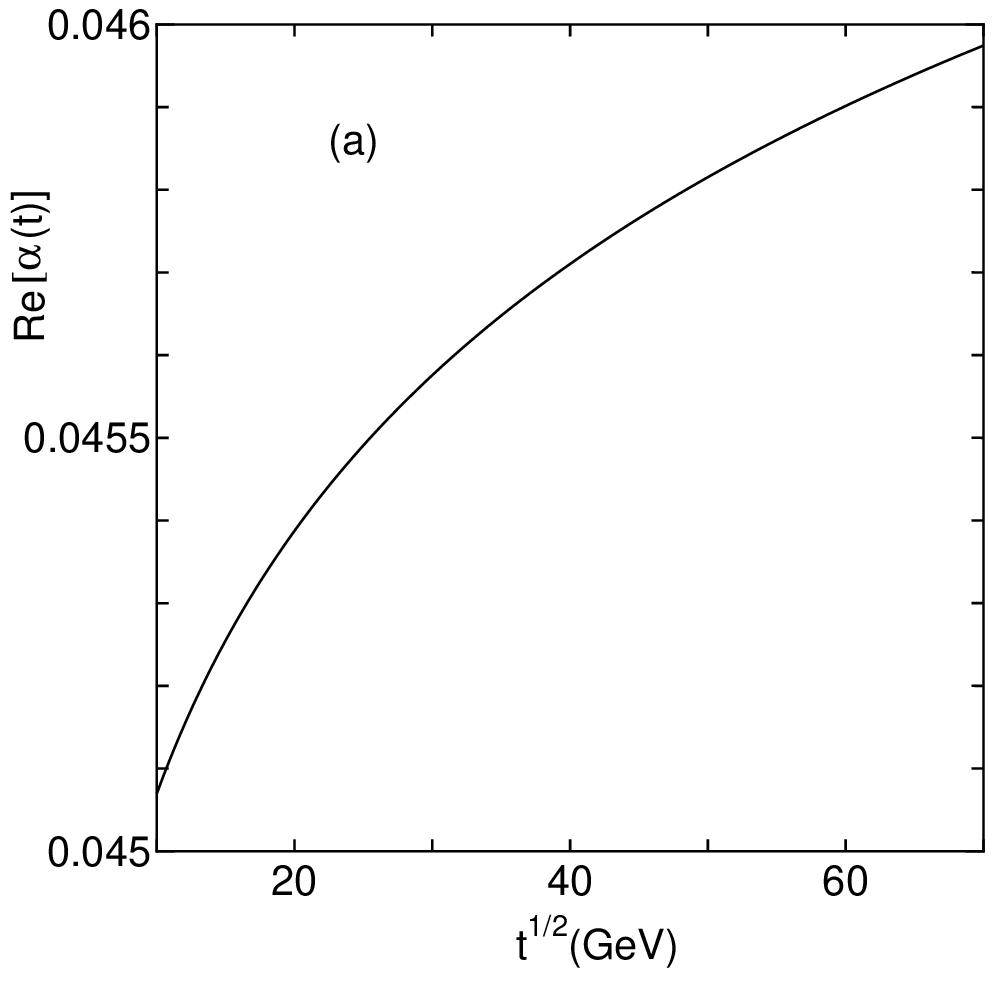}
& \includegraphics[width=.47\linewidth]{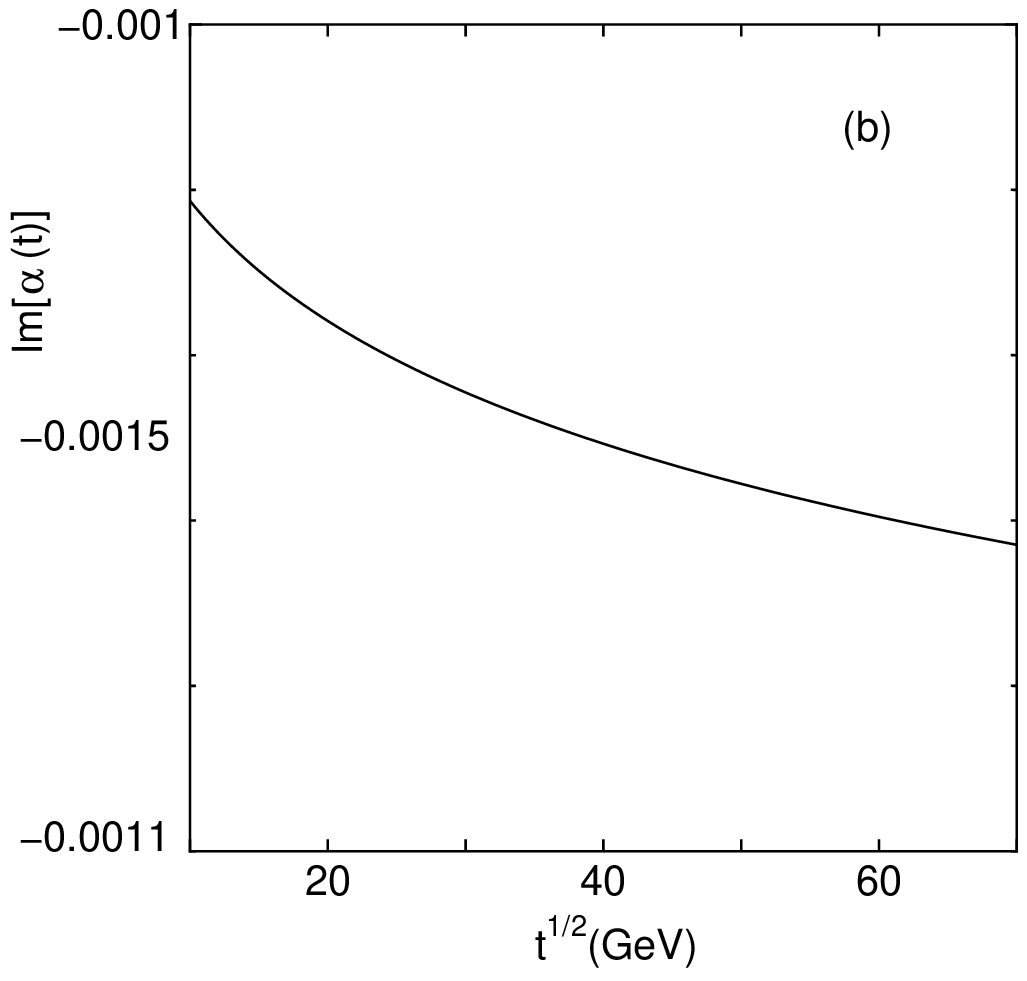}
\end{tabular}
\end{center}
Fig.1\,${\rm Re}\,\alpha$ and ${\rm Im}\,\alpha$ for $\Lambda=1.646\times
10^{47}\,{\rm GeV}$ for the time-like momentum. (a): ${\rm Re}\,\alpha(t)$. 
 (b): ${\rm Im}\,\alpha(t)$.
\end{figure}

For the space-like momentum the effective coupling constant $\alpha_S$ has a
 pole at $Q^2={Q^*}^2$ and the pole term is written as
\begin{equation}
 {\rm pole \,term\, of}\,\,\alpha_S = A^*\frac{Q^2}{Q^2-{Q^*}^2}.
                     \label{pole}
\end{equation}
Numerically, for the four loop calculation of (\ref{alpha(Q^2)}) we have 
$$
 A^*=4.50088\times 10^{-3}, \quad {Q^*}^2=a^*\Lambda^2
$$
with 
$$
  a^*=5.512275.
$$

The regularized effective coupling constant is obtained by subtracting  
the pole term (\ref{pole}), that is,
\begin{equation}
 \alpha_R(Q^2)=\alpha_S(Q^2)-A^*\frac{Q^2}{Q^2-{Q^*}^2}. \label{one pole}
\end{equation}
Although the formula is exact only for the one loop approximation, it is 
approximately correct for the higher order calculations so long as 
$Q^2 \ll \Lambda^2$. 

\section{Comparison with experiments}

We compare the effective coupling constant $\alpha_S(Q^2)$ with the experimental data \cite{okada} for the space-like and time-like momenta by taking $\Lambda$ as an 
adjustable parameter. The experimental data imply that $\alpha_S$ is an 
increasing function of $Q^2$. In Ref.\cite{okada} we have five data points for 
the space-like momentum and one for the time-like momentum. We use 
(\ref{alpha(Q^2)}), for the former and $|\alpha_S(e^{-it})|=1/\sqrt{u^2+v^2}$, 
 (\ref{time}), for the latter and determine $\Lambda$ so as to minimize 
 $\chi^2$ for the experimental data \cite{okada}. The parameter $\Lambda$ is 
 determined as follows:
$$
 \Lambda=1.646\times10^{47}\,{\rm GeV}, \,{\rm with}\,\chi^2 = 4.22 \quad
  ({\rm four\,\, loop\,\, approximation}),
$$
where the degrees of freedom is 5.

 We compare in Fig.2 (a) and (b) the calculated results with the experimental
  data of the 
 effective coupling constant for QED for the space-like and time-like 
 momentum, respectively. 
 
 \begin{figure}
 \begin{center}
 \begin{tabular}{cc}
   \includegraphics[width=.47\linewidth]{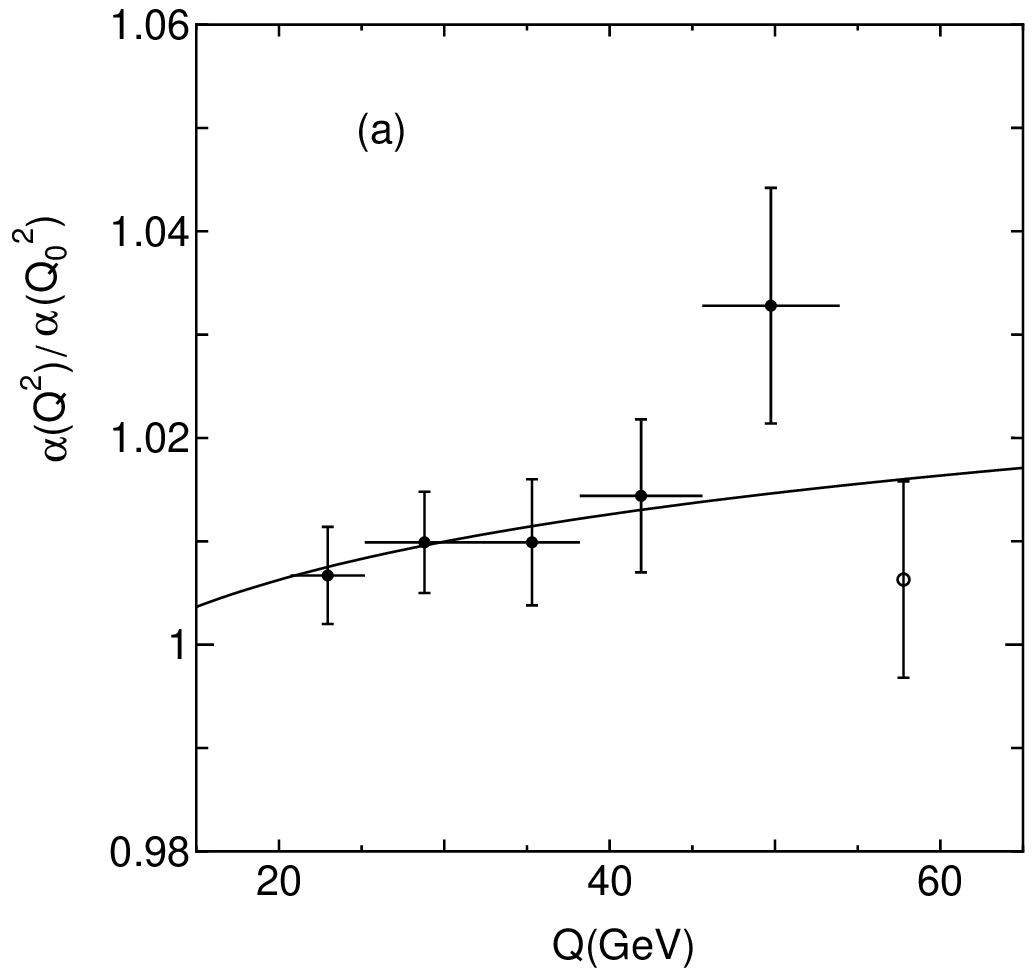}
 & \includegraphics[width=.47\linewidth]{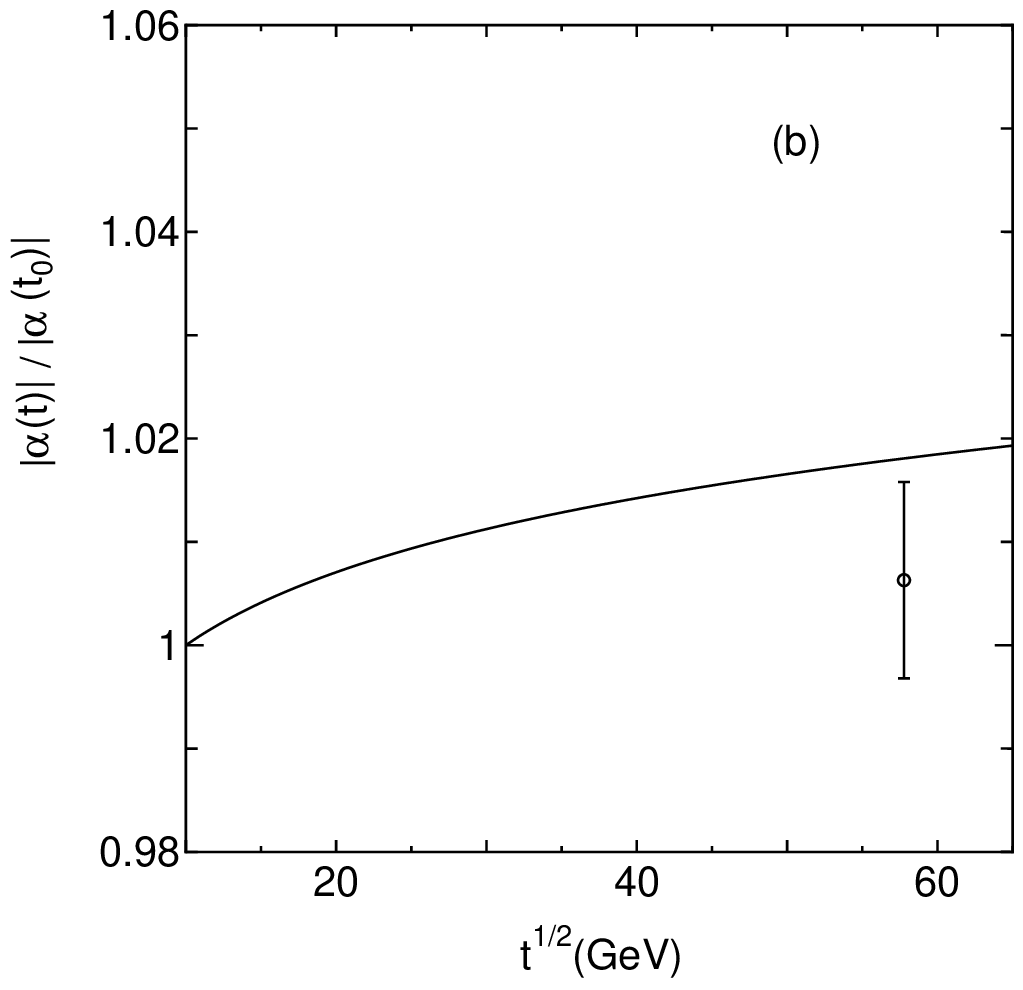}
 \end{tabular}
 \end{center}
 Fig.2 \, Effective coupling constant for QED with 
 $ \Lambda = 1.646\times10^{47}\, {\rm GeV}$. 
 (a): $\alpha(Q^2)/\alpha(Q_0^2)$ for the space-like
 momentum with $Q_0=$ 10GeV. The open circle is the data for the time-like 
 momentum, the same point as in (b).
 (b): $|\alpha(t)|/|\alpha(t_0)|$ for the time-like momentum with $\sqrt{t_0}=$ 10 GeV. Data points are taken from Re.\cite{okada}
 \end{figure}
 
Although we have used four loop calculation for the effective coupling constant, the experimental data are realized by the one loop approximation as well. 
The parameter  $\Lambda$ is then obtained to be
$$
 \Lambda = 2.166\times 10^{49}\, {\rm GeV},\, {\rm with}\, \chi^2=3.78.
 \qquad ({\rm one \,\,loop \,\,approximation}).
$$
As the value of $Q^2$ is at most 10 GeV$^2$ for the existing experiments, the
 pole term in (\ref{one pole}) can be neglected.

The value of $\Lambda$ is much larger than that of QCD for which 
$\Lambda = O(1)$ GeV. It is, however, much smaller than the mass scale  
corresponding to the Landau ghost, where 
$\Lambda^{\rm Landau}=m_e\exp(3\pi/2\alpha_r)$, with 
 $m_e$ being the electron mass and $\alpha_r$ the 
fine structure constant \cite{itzykson}. 

The large mass scale implies that some other particles than the electron 
is necessary such as Z and W or Higgs boson in the standard model. 

To conclude the paper, we remark that the regularization of the effective
 coupling constant is similar to the 
Redmond and Uretsky regularization of the Green function in QED
 \cite{redmond}.

\end{document}